\documentclass[sigconf]{acmart}

\AtBeginDocument{%
  \providecommand\BibTeX{{%
    \normalfont B\kern-0.5em{\scshape i\kern-0.25em b}\kern-0.8em\TeX}}}

\setcopyright{acmcopyright}

\begin{document}

\title{A New Dataset for Amateur Vocal Percussion Analysis}


\author{Alejandro Delgado}
\affiliation{%
  \institution{Roli Ltd.}
  \streetaddress{2, Glebe Road}
  \city{London}
  \state{England, UK}
  \postcode{E8 4BD}
}
\email{alejandro@roli.com}

\author{SKoT McDonald}
\affiliation{%
  \institution{Roli Ltd.}
  \streetaddress{2, Glebe Road}
  \city{London}
  \state{England, UK}
  \postcode{E8 4BD}}
\email{skot@roli.com}

\author{Ning Xu}
\affiliation{%
  \institution{Roli Ltd.}
  \streetaddress{2, Glebe Road}
  \city{London}
  \state{England, UK}
  \postcode{E8 4BD}}
\email{ning@roli.com}

\author{Mark Sandler}
\affiliation{%
  \institution{Queen Mary University of London}
  \streetaddress{Mile End Road}
  \city{London}
  \state{England, UK}
  \postcode{E1 4NS}}
\email{mark.sandler@qmul.ac.uk}





\renewcommand{\shortauthors}{Delgado et al.}

\begin{abstract}
The imitation of percussive instruments via the human voice is a natural way for us to communicate rhythmic ideas and, for this reason, it attracts the interest of music makers. Specifically, the automatic mapping of these vocal imitations to their emulated instruments would allow creators to realistically prototype rhythms in a faster way. The contribution of this study is two-fold. Firstly, a new Amateur Vocal Percussion (AVP) dataset is introduced to investigate how people with little or no experience in beatboxing approach the task of vocal percussion. The end-goal of this analysis is that of helping mapping algorithms to better generalise between subjects and achieve higher performances. The dataset comprises a total of 9780 utterances recorded by 28 participants with fully annotated onsets and labels (kick drum, snare drum, closed hi-hat and opened hi-hat). Lastly, we conducted baseline experiments on audio onset detection with the recorded dataset, comparing the performance of four state-of-the-art algorithms in a vocal percussion context.
\end{abstract}

\ccsdesc[500]{Applied computing~Sound and music computing}

\keywords{dataset, beatbox, vocal, imitation, percussion}

\copyrightyear{2019} 
\acmYear{2019} 
\acmConference[AM'19]{Audio Mostly}{September 18--20, 2019}{Nottingham, United Kingdom}
\acmBooktitle{Audio Mostly (AM'19), September 18--20, 2019, Nottingham, United Kingdom}
\acmPrice{15.00}
\acmDOI{10.1145/3356590.3356844}
\acmISBN{978-1-4503-7297-8/19/09}

\maketitle

\section{Introduction}
\textit{Music Information Retrieval} (MIR), which uses Digital Signal Processing (DSP) and machine learning techniques to analyse music recordings, has had a growing influence on the music industry in recent decades. Results in tasks like music genre recognition, chord estimation and source separation \cite{5} bring optimism to the field in this regard. Some of the main sub-disciplines in MIR are also receiving a significant amount of attention from independent musicians. Automatic music transcription, for instance, enables artists to learn and compose musical pieces more comfortably, while source separation helps them practice more efficiently by deriving individual instrumental tracks from the original audio mix.

In this context of innovation, a set of musical sounds worth exploring is that of \textit{vocalised percussion}. It includes vocal utterances that are articulated so to communicate a rhythmic idea, usually by imitating the sound of percussive instruments like those featured in a drum set. As such, these sounds and their dynamics could be mapped to real drum samples from a sound library so to create a realistic drum loop in seconds, making composers save time and effort prototyping rhythms without actual music knowledge.

However, despite the possibilities that these rhythmic exploration tools offer, they are seldom used today. This could be due to several factors, ranging from the limited commercial spread of the already available applications to the insufficient precision of their algorithms. As a response to the current situation, we have recorded a dataset of vocal imitations of percussion instruments, which could help researchers to shed light on the problem and head towards a reliable and easy way of creating rhythmic patterns.

A literature review of the work in vocalised percussion is presented in section 2, alongside a list of the main datasets available in the public domain. In Section 3, we introduce the Amateur Vocal Percussion (AVP) dataset, laying out its main features and detailing its production process. In section 4, we perform baseline experiments on onset detection, where we compare the performance of four well-established algorithms in the context of vocal percussion. General conclusions are drawn in section 5.

\section{Previous Work}

Between the recording of a vocal percussion performance and its mapping to a drum pattern, there are usually two main steps: onset detection and utterance classification. The former aims at localising the utterances in time, while the latter deals with the association of these with percussion instruments.

There are two main frameworks in vocal percussion analysis for classification purposes. \textit{Match and adapt} tries to find the most similar spectrum to the query one in a database of rhythmic performances \cite{19}, while \textit{onset-wise separation} divides the audio file in utterances and analyse them separately \cite{17}. Both of them use MIR analytic routines, based on frame-wise extraction and statistical aggregation of DSP descriptors (spectral centroid, zero-crossing rate, mel-frequency cepstral coefficients...) and machine learning techniques (support vector machines, decision trees...). Studies are usually carried out focusing on three classes of imitated sounds: kick drum, snare drum and hi-hat.

In most cases, a modest vocal percussion dataset is recorded for evaluation purposes, which could be oriented to general rhythmic vocal percussion \cite{17} \cite{16} \cite{2} \cite{18} or just beatboxing \cite{4} \cite{14}. Some of the mentioned work target user-specific vocal percussion analysis rather than general methods that could work for everyone, as vocal imitation styles seem to change significantly from person to person \cite{12} \cite{2}. Nevertheless, the idea of a universal model for vocal percussion classification may also be plausible, as some studies have suggested that humans, when given the task of imitating non-vocal environmental sounds, naturally give realism to these imitations by focusing on certain characteristic perceptual features of the original sound queries \cite{10}.


There are four widely adopted vocal percussion datasets in the public domain. The first one is the \textit{Beatbox dataset} by Stowell et al. \cite{14}. It gathered experienced beatboxers to record 14 audio files (one per participant) with a mean duration of 47 seconds that resulted in a total of 7460 annotated utterances. Both typical drum sounds and beatbox-specific ones were labelled, being the audio files recorded in several environmental conditions (different microphones, equipment, noise levels...). The authors also discovered that a better classification performance could be achieved by delaying the start of the first analysis frame around 23 milliseconds from the actual onset, effectively discarding the transient information of the sounds. The second one is the \textit{Live Vocalised Transcription (LVT) dataset} by Ramires et al. \cite{2}. It was recorded by 20 participants (from experienced to amateur beatboxers) using three microphones with different noise levels and sound qualities. Two files per participant are provided, one of them containing the imitation of a simple drum loop and the other a free rhythmic improvisation. A software plugin for drum transcription by vocal percussion was presented afterward \cite{8}. The third dataset by Mehrabi et al. \cite{9} is aimed at both transcribing the instruments being imitated and recognising their specific model (e.g. type of snare). In it, 14 participants with experience in music production are asked to imitate 30 percussion sounds, including different kick drums, snare drums, hi-hats, toms and cymbals. Finally, the fourth dataset is the \textit{Vocal Imitation Set} by Kim et al. \cite{400}. This is directed to vocal imitation of sound events in general, containing a total of 11242 crowd-sourced imitations of 302 different classes. It features imitations of several percussion instruments from approximately twenty participants, including opened hi-hat, snare drum and kick drum.

\section{THE AVP DATASET}

As audio recording quality is becoming better and more accessible with time, aspiring musicians are also moving away from the recording studios to a new workspace consisting of just a relatively quiet room and a laptop. Enabling independent artists to prototype rhythmic ideas without musical knowledge and in an immediate way would make the creative workflow in this setting more natural, quick and spontaneous. This is the main reason the Amateur Vocal Percussion (AVP) dataset was created.

All audio files and annotations can be found in the link \url{https://doi.org/10.5281/zenodo.3250230}

\subsection{DESCRIPTION}

The AVP dataset is characterised by the following attributes:

\begin{itemize}
\item 28 participants
\item A total of 9780 utterances in 280 audio files
\item Annotated onsets and labels (kick, snare, closed hi-hat and opened hi-hat)
\item Recorded with one microphone (MacBook Pro's built-in mic)
\item 2 modalities: personal imitations and fixed imitations
\item 5 files for each modality each (four files with utterances of the same class and one with an improvisation)
\end{itemize}

In contrast with the previously cited datasets, the AVP focuses exclusively on people with little or no experience in beatboxing, making it suitable for the goal mentioned before as well as functioning as an extension of past datasets. It is also directed to query-by-example applications, allowing algorithms to be trained on several isolated example utterances (around twenty-five per label) before they are tested on the improvisation files.

The AVP is the largest vocal percussion dataset both in number of participants and number of utterances. It also incorporates a second subset of fixed drum imitations, i.e., four specific sounds given to all participants to imitate each one of the four instruments. This subset was recorded in order to investigate how practical a fixed technique of performing vocal percussion could be, as deriving accurate algorithms to classify the personal subset as a whole is relatively challenging. Phoneme recognition systems, for instance, are likely to help in this particular endeavour; though, in comparison with regular speech, different articulations and expressive dynamics are likely to be encountered in rhythmic vocal percussion.

\begin{table}
  \caption{Dataset content summary (utterances)}
  \label{tab:freq}
  \begin{tabular}{lccc}
    \toprule
    Instrument - `\textit{label}' & Personal & Fixed & Improvisation\\
    \midrule
    Kick Drum - `\textit{kd}' &  \textit{799}      & \textit{818}       & \textit{1201}\\
    Snare Drum - `\textit{sd}' & \textit{813}       & \textit{839}        &  \textit{811}\\
    Closed Hi-Hat - `\textit{hhc}' & \textit{799}      & \textit{833}        & \textit{673}\\
    Opened Hi-Hat - `\textit{hho}' & \textit{816}   & \textit{830}        & \textit{548}\\
    \bottomrule
\end{tabular}
\end{table}

The main weakness of the AVP dataset is that, due to the lack of available resources at the time, it was recorded without using any extra microphones. However, when it comes to the dataset's goal, this particularity does not affect it in a critical way, as it is directed to drum pattern query by vocal percussion in an indoor setting. Also, the realism of the recording context, featuring quiet environmental noises as well, could be proven helpful for the algorithms to perform well in non-controllable external conditions.

The total number of vocal percussion utterances belonging to each of the four classes are gathered in table 1.

\subsection{RECORDING METHODOLOGY}

The materials used to record the dataset were a MacBook Pro laptop, GarageBand software and a closed room of approximately 40 m$^3$. The experiment took an average of 15 minutes per participant. 

The first step of the process was a small survey about the role that music played in the participant's life. This informal one-minute questionnaire was oriented towards relaxing the participants and get them ready for the task. Right after this, a standard loop featuring kick and snare was presented and the participant was asked to both reproduce it vocally and write down the onomatopoeias of the sounds he/she used in a notebook page. This performance was intended as a first contact with the process and it did not get recorded, while the annotation of the onomatopoeias was done to facilitate the recalling of imitations and to better stick to them. As it turned out to be hard for participants to imitate complex beats with hi-hat included in the preliminary tests, five isolated utterances of both closed hi-hat and open hi-hat sounds were presented instead. The participants decided their imitations and wrote down their onomatopoeias in the same way as with the kick drum and snare drum.

Once the participants were familiarised with the task, they were asked to sit naturally in front of the computer, as they would usually do, and the recording of the dataset started. There were two recording modalities taking place: personal and fixed. For the personal modality, participants used their own vocal imitations to record around twenty-five utterances of kick drum, snare drum, closed hi-hat and opened hi-hat sounds in four separated audio files. The utterances followed a simple rhythmic loop (one crotchet and two quavers) and a 90-BPM metronome track was provided through headphones to the participants in order to record them. An audio file of improvised vocal percussion featuring all or most instruments was recorded afterward. For the fixed modality, the procedure above was repeated once more, but now four specific sounds were given to the subjects. These fixed imitations were based on speech syllables so to feel natural for participants to reproduce, and their timbral characteristics were intended to mimic the imitated percussion instruments. A ``pm'' syllable would correspond to the kick drum, ``ta'' to the snare drum, ``ti'' to the closed hi-hat and a ``chi'' to the opened hi-hat. The articulation of these utterances, as they were performed in a percussive setting, generally resulted in brighter transient signals and more inharmonic steady state signals compared to usual speech utterances.

\subsection{CLEANING AND ANNOTATION}

Once the raw audio files were recorded, two post-processing stages took place to prepare them for analysis: trim silent regions at the beginning and the end of each file to reduce their size and remove in-between passages where the participants made accidental mistakes or notable pauses. The annotation of the files was manually carried out by the author right after this cleaning process, using Sonic Visualiser \cite{22} to write down both onset locations and class labels. As illustrated in table \ref{tab:freq}, the tag `\textit{kd}' was used for kick drum, `\textit{sd}' for snare drum, `\textit{hhc}' for closed hi-hat and `\textit{hho}' for opened hi-hat.

The choice of the exact starting point of a sound event is generally considered to be dependent on the task at hand and, thus, a matter of convenience. For instance, if the goal is preparing an utterance with a long transient to be used as a sound triggered by a MIDI message, the onset would sometimes be preferably placed near the point of maximum energy in the signal. In our case, we decided to place the onsets at the very beginning of the utterance, where the percussive transient starts to build the sound. This is due to the fact that our primary goal is to classify the utterance, and its transient region could be informative as well.

An important note to make regarding the annotation process is that, in a few occasions, two utterances were very close to each other or even appeared to merge in one sound. A joint approach of waveform visualisation, spectrogram visualisation and listening at quarter speed was employed to solve the ambiguity in these cases. An example of how these specific annotations were approached is illustrated in figure 1.

\begin{figure}
\centering
\includegraphics[scale=0.3]{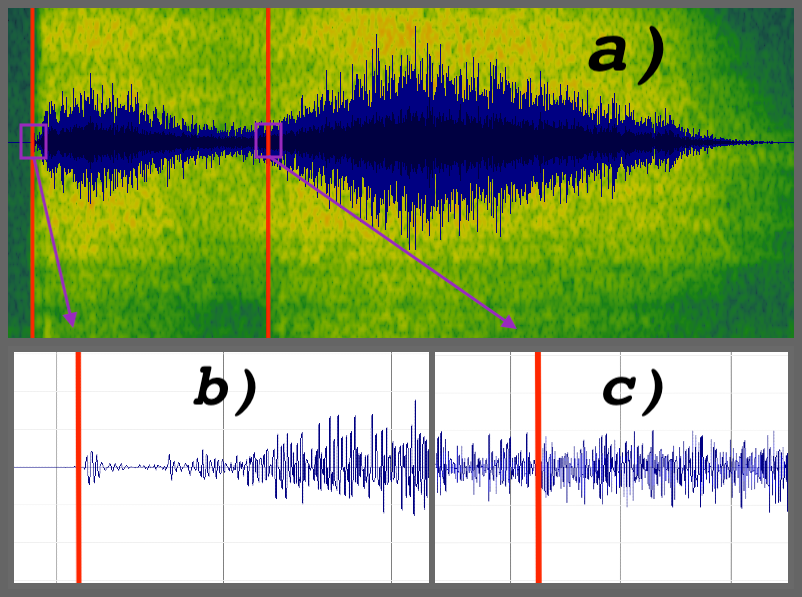}
\caption{Annotation of two onsets, corresponding to two /s/ phonemes. Figure 1a) displays the utterances with their onsets marked in red. The waveform is plotted in blue and the magnitude spectrogram in green. Figures 1b) and 1c) are close captions of the first and the second onset respectively.}
\end{figure}

\subsection{FURTHER OBSERVATIONS}

A list of observations worth commenting on was made while recording and listening back the audio files in the AVP dataset. Some of them are the following:

\begin{itemize}
    \item A small set of recorded utterances exhibit a form of double percussion (such as ``suc'' or ``brrr''), making them challenging for the onset detector to output the events as a whole without splitting them.
    \item Various participants got mistaken in few occasions when recording the improvisation files, using other speech-like phonemes like ``cha'' instead of ``ta'' when imitating the snare drum or ``dm'' instead of ``pm'' when imitating the kick drum. These passages were omitted in the final version of the dataset.
    \item Several participants reported that the given fixed sounds, despite their resemblance with the original drum sounds, felt unnatural for them to perform.
    \item Participants generally improvised non-complex and predictable loops, which could make the classification routines benefit from a rhythmic pattern analyser.
\end{itemize}

Finally, there have been three participants whose utterances within the personal dataset were either not consistent with each other, unintelligible or practically indistinguishable from the rest. The audio files and annotations pertaining to these cases are stored in the ``Discarded" folder. Despite this, they could still be used for classification purposes as long as the improvisation files, where the ambiguities occur, are excluded.

\section{EXPERIMENT ON ONSET DETECTION}

The problem of onset detection is a recurrent one in MIR literature. Most of its subfields, in one way or another, need to face it at some point when trying to make music analysis fully automatic. Vocal percussion resembles speech when it comes to the articulation of utterances, making the job slightly more challenging than with regular percussion, which exhibits short and well-defined attacks. All vocal percussion utterances are usually composed by one fricative phoneme that is sometimes followed by a vowel phoneme. This is illustrated in figure 2.

\begin{figure}
\centering
\includegraphics[scale=0.40]{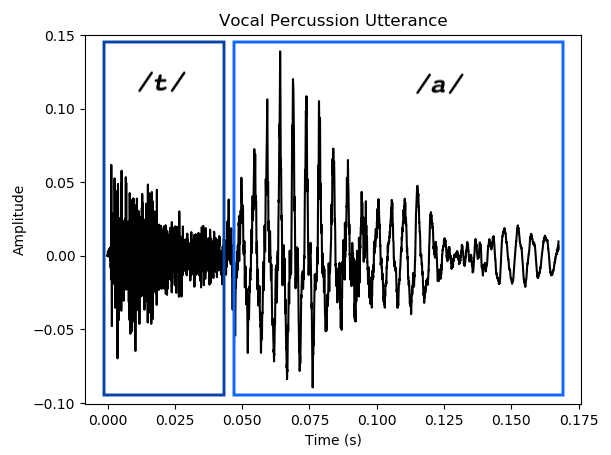}
\caption{Piece of waveform pertaining to a single vocal percussion utterance. One can appreciate the plosive phoneme /t/ and the vowel phoneme /a/ composing the ``ta'' syllable as distinct components in the audio file.}
\end{figure}

Here we describe a baseline study on onset detection for vocal percussion using the AVP dataset. It aims at localising the time instances when the utterances begin up to a tolerance margin of 50 ms \cite{5}. This section is merely intended to be a brief analysis of the task, provoking further discussions and laying down foundations for future work.

\subsection{MODELS}


In this section, we present the onset detection methods that will be evaluated in this study. We will first do a quick overview of the \textit{spectral flux} descriptor and its relevance to the task. Then, we will take a look at four state-of-the-art algorithms for audio onset detection. The first two, \textit{Convolutional Neural Networks (CNN)} and \textit{Recurrent Neural Networks RNN}, are based on deep machine learning methods while the other two, \textit{High-Frequency Content (HFC)} and \textit{Complex}, are based on traditional signal processing routines.

\subsubsection{Spectral Flux (for onset refinement)}

Despite this descriptor will not be used for onset detection but rather to refine the locations of the detected onsets, all four algorithms investigated here are either based on or inspired by it. For this reason, we will describe it before we introduce the rest of the methods, considering it a theoretical reference point.

The spectral flux is a frame-wise measure of how quickly the magnitude in each frequency bin of the spectrum changes over time \cite{15}. The sharper the change in a particular region, the more probable a percussive onset is occurring there. If $x$ is the audio waveform and $X$ is its Fourier transform, the spectral flux is expressed in the following way:

\begin{equation}
  S F(n)=\sum_{k=-\frac{N}{2}}^{\frac{N}{2}-1} H(|X(n, k)|-|X(n-1, k)|)
\end{equation}

where $k$ is a frequency bin of the $n$th frame and
\begin{math}
  H(x)=\frac{x+|x|}{2}
\end{math}
is called the half-wave rectifier function, which restricts the spectral flux to only output positive changes in time, i.e., when the flux increases. The higher a peak in the spectral flux function, the more likely an onset is taking place in that location.

\subsubsection{Convolutional Neural Network (CNN)}

CNNs \cite{23} are feed-forward artificial neural networks composed of convolutional layers. The neurons in these layers compose a set of small local filter kernels to analyse the input, creating several feature maps as a result. A convolutional layer could be followed by a pooling layer, which subsamples these resulting feature maps following a certain set of rules, or by a fully connected layer to introduce non-linearities in the process. These last layers are specially useful for classification purposes.

CNNs are especially popular in computer vision, and they achieve high accuracies when detecting edges in images. The main idea behind their use for audio onset detection is that they can effectively detect edges in the magnitude spectrograms as well, which usually correspond to audio onsets. In this way, the usual input for the CNN is composed by three 80-band mel spectrograms with constant hop size but different window sizes so to have both high frequency and time resolution. The details of the original network's architecture can be found in \cite{1}. It was discovered in the same study that the CNN, like spectral flux based techniques, computes spectral differences over time and uses them to estimate the likelihood of an onset happening in a certain time region. The implementation of the CNN-based onset detection method, which this study follows, is contained in Madmom's \textit{CNNOnsetProcessor} \cite{11}.

\subsubsection{Recurrent Neural Network (RNN)}

The functioning of an RNN, in contrast with CNNs, is dependent on the information acquired from past inputs, allowing these networks to model complex data sequences in a non-linear way. Long Short-Term Memory (LSTM) units in RNNs \cite{24} allow the networks to better handle dependencies in time and avoid the exploding/vanishing gradient problem.

When spectrograms are fed to RNNs, these networks can detect and model changes in time by contrasting observations from current frames with information from previous ones. Their routines, then, would be analogous to those of the spectral flux onset detector, with RNNs taking information from more past frames into account. The original RNN model for onset detection can be found in \cite{7}. This time, the authors use three Bark spectrograms with constant hop size and different window sizes as inputs, and the resulting model is also optimised to handle less data than other onset detection algorithms. This makes it suitable for real-time onset detection, using unidirectional (causal) RNNs without LSTM cells. The implementation of the RNN-based onset detection method, which this study follows, is contained in Madmom's \textit{RNNOnsetProcessor} \cite{11}. We use the non-real-time version, with bidirectional RNNs.

\subsubsection{High Frequency Content Method (HFC)}

In general, most of the energy of a sound's steady state part is located in the lower frequencies, being the higher frequencies more represented in its transient part. When a percussive onset occurs, the sudden increase in energy tends to be particularly more pronounced at higher frequencies, indicating the beginning of its transient.

The HFC method for onset detection \cite{3} exploits this insight making use of the high-frequency content descriptor, i.e., the summation of the bin magnitudes of the spectrum multiplied by their own bin indices. This can be formalised in the following way:

\begin{equation}
  \mathrm{HFC}=\sum_{i=0}^{N-1} i|X(i)|
\end{equation}

for a STFT spectrum $X$ with bin indices $[0...n]$. The result, which is linearly biased towards high frequencies, is then fed to a detection function that returns the predicted onset location. This detection function is based on both the high frequency energy flux between the two adjacent frames and the normalised high frequency content of the current frame. The implementation that this study follows is included in Aubio's \textit{AubioOnset} \cite{11}. It has also been reported that the HFC method works especially well for vocal percussion \cite{2}.

\subsubsection{Complex Method (Complex)}

The phases of the partials in the steady state of a sound can be accurately predicted from past frames, as their frequencies and amplitudes remain constant in this region. In the case of its transient part, however, these phases tend to evolve in a non-linear, unpredictable way. We can use this particularity of the transient's STFT phase spectrum to locate it within the waveform.

The complex method for onset detection effectively combines both energy-based and phase-based approaches simultaneously by detecting onsets in the complex domain \cite{20}. The complex detection function is then pre-processed with a weighted moving average and fed to a peak-picking algorithm that outputs the predicted onset locations. The implementation that this study follows is included in Aubio's \textit{AubioOnset} \cite{11}.

\subsection{EVALUATION AND RESULTS}

We carry out three different studies to retrieve the parameters that produce the best performances for each algorithm in a vocal percussion context. Grid search is performed for each parameter using ten linearly spaced values. We test the algorithms on all the utterances in the AVP dataset and the accuracy of each performance is measured using the F1-Score.






The first analysis aims at retrieving the best values for the method-specific parameters. These parameters are the \textit{peak picking threshold} of the onset probability functions for all methods and the \textit{frame size} for the HFC and Complex ones. It has been found that the best results for each algorithm are reached by defining a threshold of 0.55 for CNN, 0.30 for RNN, 0.8 for HFC and 0.7 for complex. A frame size of 11 ms gave the best results for both the HFC and the Complex methods.

The second study explores how a \textit{minimum separation} between onsets could further benefit the algorithms' performance. This means that in case two of the predicted onsets happen to be separated in time by less than the value of a certain threshold, the second onset will be automatically discarded. This appears to be relevant to our case for two reasons. The first one is that two vocal percussion utterances occur seldom or never simultaneously, i.e., we are dealing with a monophonic transcription problem. Therefore, there will always exist a minimum separation between onsets. The second reason is that these algorithms can easily confuse the transient of a vowel phoneme with the onset of a new utterance, lowering their precision score significantly. This inconvenience could be avoided with the inclusion of a fixed minimum separation between onsets, as the onset from the vowel phoneme could be effectively discarded.

Results for this second study are displayed in figure 3. The main observation is that both the CNN and RNN methods benefit from the inclusion of a minimum onset separation, reaching the peak of their performance at the 90 ms mark, whilst the HFC and the Complex methods show a detriment in performance when this separation is applied. A reason for this phenomenon could be that both the HFC and the Complex algorithms are based on distinctive features of percussive transients, being them the high-frequency content and the predictability of the phase spectrum respectively. For these methods, the peaks in the onset probability function corresponding to plosive phonemes (see figure 2) would be appreciably higher than the peaks relative to the vowel phonemes. Hence, setting an appropriate value for the peak-picking threshold would be enough to separate these types of phonemes, and a minimum onset separation would only make the algorithm discard correct onsets in situations when two real onsets are significantly close to each other. Deep learning based algorithms, on the other hand, are optimised so to detect musical onsets in general and do not explicitly rely on these characteristics of percussive transients, which makes them more likely to detect the onsets of the vocal phonemes as well. Thus, setting a minimum onset separation is a simple yet effective way for them to ignore onsets coming from vowel phonemes, at the expense of discarding correct onsets that are close in time.

\begin{figure}
\centering
\includegraphics[scale=0.40]{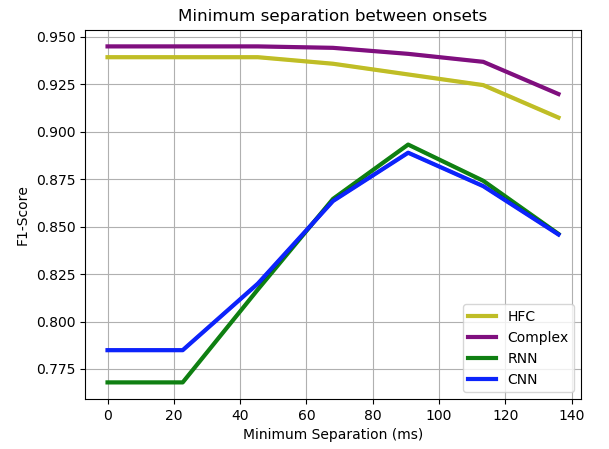}
\caption{Effect of different minimum separation values on the performance of the four onset detection algorithms.}
\end{figure}

Finally, we attempt to \textit{refine} these onsets in time using the spectral flux descriptor. The goal here is to relocate each predicted onset in a neighbouring point which is closer to the real onset. More technically, it consists in minimising the mean absolute deviation of the predicted onsets with respect to the real onsets without significantly affecting the performance (F1-Score). We approach this task by centring an analysis window in the predicted onset and then finding the maximum of the spectral flux function in that region, where the real onset is more likely to be. We optimised the length of this analysis window selecting the value that achieved the best balance in the performance vs. precision trade-off. Specifically, we set a maximum allowed drop in performance of 1\% of its original F1-Score, considering only the parameter values that did not make the algorithms exceed that limit.

The final results are gathered in table 2. The HFC and the Complex methods perform the best when it comes to accuracy and the onsets derived from deep learning methods are the closest in time to the real onsets. Results also present CNNs and RNNs as slow compared to HFC and Complex, which are 38 and 27 times faster respectively.


\begin{table}
  \label{tab:freq2}
  \begin{tabular}{lccc}
    \toprule
    Method & F1-Score & Deviation & Duration\\
    \midrule
     \textbf{CNN} & 0.89       & \textbf{9} ms  & 267 s\\
     \textbf{RNN} & 0.89 & 10 ms   & 266 s\\
     \textbf{HFC} & 0.94      & 17 ms   & \textbf{7} s\\
     \textbf{Complex} & \textbf{0.95}   & 21 ms   & 10 s\\
     \midrule
     \textbf{CNN-SF} & 0.89       & \textbf{6} ms  & -\\
     \textbf{RNN-SF} & 0.88 & 9 ms   & -\\
     \textbf{HFC-SF} & 0.93      & 11 ms   & -\\
     \textbf{Complex-SF} & \textbf{0.94}   & 11 ms   & -\\
    \bottomrule
\end{tabular}
\caption{Final results of the onset detection study for vocal percussion. The '-SF' suffix indicates the inclusion of spectral flux refinement in the process. From left to right, it is displayed the F1-Score, the mean absolute deviation of the predicted onsets with respect to the real ones, and the duration of the analysis for all onsets using the same laptop.}
\end{table}

\section{CONCLUSION}

In this piece of work, we have presented the AVP dataset, comprising 9780 utterances of vocal percussion recorded by 28 participants. This aims at improving algorithms for drum pattern query by vocal imitation so that independent musicians can accelerate their creative routines by sketching realistic percussion rhythms on the fly. Baseline experiments on utterance onset detection were carried out in the final section. It was shown that methods relying on specific DSP features outperformed deep learning based techniques in the context of vocal percussion, laying the groundwork for novel approaches to improve upon.

\begin{acks}

$
\begin{array}{l}
\includegraphics[scale=0.1]{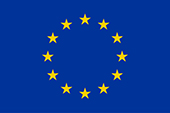}
\end{array}
$ This project has received funding from the European Union's Horizon 2020 research and innovation programme under the Marie Sk$\l{}$odowska-Curie grant agreement No. 765068

A warm and special thank you to the people that participated in the creation of this dataset.
\end{acks}

\bibliographystyle{ACM-Reference-Format}
\bibliography{Bibliography_AM}

\end{document}